\documentstyle[preprint,prd,aps]{revtex}
\begin{document}
\draft
\pagenumbering{roma}
\author{Yuan-Ben Dai$^a$, Chao-Shang Huang$^a$, Ming-Qiu Huang$^{b,a}$}
\address{$^a$ Institute of Theoretical Physics, Academia Sinica, P.O.Box 2735, Beijing 100080, China}
\address{$^b$ CCAST (World Laboratory) P.O. Box 8730, Beijing, 100080}
\title{${\cal O}(1/m_Q)$ Order Corrections to Masses of Excited Heavy Mesons From QCD Sum Rules}
\date{\today}
\maketitle
\thispagestyle{empty}
\vspace{15mm}
\begin{abstract}
The ${\cal O}(1/m_Q)$ corrections to masses of excited heavy mesons are studied with sum rules in the heavy 
quark effective theory. Numerical results for the matrix elements of the heavy quark kinetic energy operator $\cal K$ and chromomagnetic interaction operator $\cal S$ are obtained for the lowest excited doublets $(0^+, 1^+)$ and $(1^+, 2^+)$.
\end{abstract}
\vspace{4mm}
\newpage
\pagenumbering{arabic}

\section{Introduction}
\pagenumbering{arabic}
In recent years there has been a continuous interest in the study of excited
 heavy mesons composed of a heavy quark and a light anti-quark. This interest arises from several reasons. Some of these excited states have been observed in experiments. They will be objects of further study in 
future experiments with B-factories. In particular, they are useful for
 tagging in CP experiments \cite{eichten}. Theoretically, the relative simplicity of the dynamical condition makes it of interest for exploring the internal dynamics of systems containing a light quark.

Important progresses in the theoretical description of such systems have 
been achieved with the development of the Heavy Quark Effective Theory (HQET) \cite{grinstein}. Based on the spin-flavor symmetry of QCD, valid in the 
limit of infinite heavy quark mass $m_Q$, this framework provides a 
systematic expansion of heavy hadron spectra and transition amplitudes in 
terms of the leading contribution, plus corrections decreasing as powers of $1/m_Q$. However,  to
obtain some detailed predictions one needs to combine it with some non-perturbative methods.
The spectra and decay widths of heavy meson excited states have been studied with the $1/m_Q$ expansion in the relativistic Bethe-Salpeter equations in \cite{dai1,dai2}. They can also be studied with the QCD sum rules in HQET which have been used both for ground states of heavy mesons \cite{bagen,ball,neubert} and for lowest excited heavy meson doublets $(0^+, 1^+)$ and $(1^+, 2^+)$ at leading order of the $1/m_Q$ expansion\cite{huang}. For related results, see \cite{colangelo}. 

 The effective Lagrangian of the HQET, up to order $1/m_Q$, can be written as
\begin{equation}
\label{Leff}
   {\cal L}_{\rm eff} = \bar h_v\,i v\!\cdot\!D\,h_v
   + \frac{1}{2 m_Q}\,{\cal K}
   + \frac{1}{2 m_Q}\,{\cal S}+{\cal O}(1/m_Q^2) \,,
\end{equation}
where $h_v(x)$ is the velocity-dependent field related to the
original heavy-quark field $Q(x)$ by
\begin{equation}
   h_v(x) = e^{i m_Q v\cdot x}\,\frac{1+\rlap/v}{2}\,Q(x)\;,
\end{equation}
${\cal K}$ is the operator of nonrelativistic kinetic energy with a negative sign defined as
\begin{equation}
\label{kinetic}
{\cal K}=\bar h_v\,(i D_\perp)^2 h_v\;,
\end{equation}
where $D_\perp^\mu=D^\mu-(v\cdot D)\,v^\mu$, with $D^\mu=\partial^\mu-i g\, A^\mu$ is the gauge-covariant derivative, and $\cal S$ is the Pauli term, describing the chromomagnetic interaction:
\begin{equation}
\label{pauli}
{\cal S}=\frac{g}{2}\,C_{mag}(m_Q/\mu)\;
   \bar h_v\,\sigma_{\mu\nu} G^{\mu\nu} h_v\;,
\end{equation}
where $C_{mag}=\displaystyle{\left(\alpha_s(m_Q)\over \alpha_s(\mu)\right)^{3/{\beta}_0}}$, ${\beta}_0=11-2n_f/3$. Apart from leading contribution, the Lagrangian density contains to ${\cal O}(1/m_Q)$ accuracy two additional  operators $\cal K$ and $\cal S$. The matrix elements of these two operators over heavy meson states are fundamental parameters in the $1/m_Q$ expansion with HQET that should 
be either determined phenomenologically from experimental data or estimated using a nonperturbative theoretical approach. The matrix element of the chromomagnetic interaction operator is the leading contribution to the mass
splitting among the states belonging to the same heavy meson doublet. 
 
In our previous work \cite{huang},  the masses of lowest excited heavy 
meson doublets $(0^+,1^+)$ and $(1^+,2^+)$ are calculated at the leading order of $1/m_Q$ expansion by using the QCD sum rule approach in HQET. As an extension of this work, in the present article we shall use the QCD sum rules in HQET to obtain quantitative estimates of the matrix elements of the kinetic energy and chromomagnetic interaction operators, which are responsible for the kinetic energy and the chromomagnetic mass corrections. 

The remainder of this paper is organized as follows. In Section \ref{sec2} we begin with a brief review on the interpolating currents for excited heavy mesons  and some important properties of these currents. Some results of two-point functions for the doublets $(0^+,1^+)$ and $(1^+,2^+)$ at the leading order of the $1/m_Q$ expansion are outlined. Section \ref{sec3} is devoted to ${\cal O}(1/m_Q)$ corrections, the sum rules for the relevant matrix elements of kinetic energy and chromomagnetic interaction operator are derived.  Finally, numerical results are presented in Section \ref{sec4}.


\section{Interpolating currents for heavy mesons of arbitrary spin and parity
and two-point correlation function}
\label{sec2}
A basic element in the application of QCD sum rules to excited heavy mesons is 
to choose a set of appropriate interpolating currents  in terms of quark fields which create 
(annihilate) definite states of excited heavy mesons. For excited heavy mesons with arbitrary spin and parity this problem has been studied in our previous paper \cite{huang}. Here we briefly outline the results of this analysis which
will be used in this article. 

From the general form of the leading order B-S wave function obtained in \cite{dai3}, we can write the general expression of the interpolating 
current creating a excited heavy meson with arbitrary spin $j$ and parity $P$ as
\begin{mathletters}
\label{current}
\begin{eqnarray}
\label{current1}
J_{j,P,i}^{\dag\alpha_1\cdots\alpha_j}(x)=\bar h_v(x)\Gamma_{j,P,i}^{\{\alpha_1\cdots\alpha_j\}}({\cal {D}}_{x_t})q(x)\;,
\end{eqnarray}
or
\begin{eqnarray}
\label{current2}
J_{j,P,i}^{'\dag\alpha_1\cdots\alpha_j}(x)=\bar h_v(x)\Gamma_{j,P,i}^{
\{\alpha_1\cdots\alpha_j\}}({\cal {D}}_{x_t})(-i)\not\!{\cal D}_{x_t}q(x)\;,
\end{eqnarray}
\end{mathletters}
where $i=1,2$ correspond to two serieses of doublets of the spin-parity 
$[j^{(-1)^{j+1}},(j+1)^{(-1)^{j+1}}]$ and $[j^{(-1)^{j}},(j+1)^{(-1)^{j}}]$ respectively, ${\cal {D}}_{x_t}$ is the covariant derivative $\displaystyle{\frac{{\partial}}{\partial x_t}-igA_t(x)}$, subscript $t$
denotes the transverse component of a 4-vector
and
\begin{eqnarray}
\label{gamma}
\Gamma^{\{\alpha_1\cdots\alpha_j\}}({\cal {D}}_{x_t})=\text{Symmetrize}\left\{\Gamma^{\alpha_1\cdots\alpha_j}({\cal {D}}_{x_t})-
\frac{1}{3}g_t^{\alpha_1\alpha_2}g_{\alpha_1^{'}\alpha_2^{'}}^t
\Gamma^{\alpha_1^{'}\alpha_2^{'}\alpha_3\cdots\alpha_j}({\cal {D}}_{x_t})\right\}\nonumber\;
\end{eqnarray}
with $g_t^{\alpha_1 \alpha_2}=g^{\alpha_1\alpha_2}-v^{\alpha_1}v^{\alpha_2}$. 
For the doublets of the spin-parity 
$[j^{(-1)^{j+1}},(j+1)^{(-1)^{j+1}}]$ and $[j^{(-1)^{j}},(j+1)^{(-1)^{j}}]$, the  expressions for $\Gamma^{\alpha_1\cdots\alpha_j}({\cal {D}}_{x_t})$
have been explicitly given in \cite{huang,dai3} as  
\begin{eqnarray}
&&\Gamma({\cal {D}}_{x_t})=\left\{
\begin{array}{l}\label{gamma1}
\displaystyle{\sqrt{\frac{2j+1}{2j+2}}}\gamma^5(-i)^j{\cal {D}}_{x_t}^{\alpha_2}\cdots{\cal {D}}_{x_t}^{\alpha_j}({\cal {D}}_{x_t}^{\alpha_1}-{\frac{j}{2j+1}}\gamma_t^{\alpha_1}\not\!{\cal {D}}_{x_t})~~\:\;\mbox{for}\; j^{(-1)^{j+1}}\;,\\[4mm]
\displaystyle{\frac{1}{\sqrt{2}}}\;\gamma_t^{\alpha_1}(-i)^j{\cal {D}}_{x_t}^{\alpha_2}\cdots{\cal {D}}_{x_t}^{\alpha_{j+1}}\hspace{3.6cm}\:\;\mbox{for}\; (j+1)^{(-1)^{j+1}}\;,
\end{array}\right.\\[2mm]\nonumber &&\\
&&\Gamma({\cal {D}}_{x_t})=\left\{
\begin{array}{l}\label{gamma2}
\displaystyle{\frac{1}{\sqrt{2}}}\gamma^5\;\gamma_t^{\alpha_1}(-i)^j{\cal {D}}_{x_t}^{\alpha_2}\cdots{\cal {D}}_{x_t}^{\alpha_{j+1}}\hspace{3.5cm}\:\;\mbox{for}\; (j+1)^{(-1)^{j}}\;,\\[4mm]
\displaystyle{\sqrt{\frac{2j+1}{2j+2}}}(-i)^j{\cal {D}}_{x_t}^{\alpha_2}\cdots{\cal {D}}_{x_t}^{\alpha_j}({\cal {D}}_{x_t}^{\alpha_1}-{\frac{j}{2j+1}}\gamma_t^{\alpha_1}\not\!{\cal {D}}_{x_t})~~~~~~~\:\;\mbox{for}\; j^{(-1)^{j}}\;.
\end{array}\right.
\end{eqnarray}
The currents hermitian conjugate to (\ref{current1}) and (\ref{current2})
\begin{mathletters}
\label{conju}
\begin{eqnarray}
\label{conju1}
J_{j,P,i}^{\alpha_1\cdots\alpha_j}(x)&=&\bar q(x)\bar{\Gamma}_{j,P,i}^
{\{\alpha_1\cdots\alpha_j\}}(\stackrel{\leftarrow}{\cal {D}}_{x_t})h_{v(x)}\;,\\[2mm]
\label{conju2}
J_{j,P,i}^{'\alpha_1\cdots\alpha_j}(x)&=&\bar q(x)(i\stackrel{\leftarrow}
{\not\!{\cal D}}_{x_t})\bar{\Gamma}_{j,P,i}^{\{\alpha_1\cdots\alpha_j\}}(\stackrel{\leftarrow}{\cal {D}}_{x_t})h_{v(x)}\;,
\end{eqnarray}
\end{mathletters}
where $\bar\Gamma=\gamma^0\Gamma^{\dag}\gamma^0$, correspond to the annihilation of the same meson.

The currents defined in (\ref{current}) and (\ref{conju}) have nice properties. 
We had shown in \cite{huang} that two currents in the sets (\ref{current1}) and (\ref{current2}) with not identical values of $j$, $P$, $i$ never mix in the $m_Q\to\infty$ limit either in the heavy meson or in the quark-gluon level and heavy quark symmetry is explicit with these currents. This verifies that they are the appropriate interpolating currents for heavy meson states with definite $j$, $P$ and the
light quark angular momentum $j_l$ which is conserved in QCD in $m_Q\to\infty$ limit.

The properties for currents mentioned above are important for applications to QCD sum rules for excited heavy mesons. If one use other currents or study the sum rules in full QCD, there are in general contributions from two nearby poles corresponding to states of the
same $j$, $P$ which have different values of the total angular momentum of the light component $\displaystyle{j_l=j\pm\frac{1}{2}}$ in the $m_Q\to\infty$ limit. Their contributions may not be separated correctly. Furthermore, the mixing
of such two states can be calculated within our formalism by introducing the
 ${\cal O}(1/m_Q)$ terms in the Lagrangian in HQET. 

Let $|j,P,i\rangle$ be a heavy meson state with the quantum numbers $j$, $P$, $i$ in the $m_Q\to\infty$ limit, together with the corresponding interpolating current, we have \cite{huang}
\begin{eqnarray}
\label{decay}
\langle 0|J_{j,P,i}^{\alpha_1\cdots\alpha_j}(0)|j',P',i'\rangle=f_{Pj_l}\delta_{jj'}\delta_{PP'}\delta_{ii'}\eta^{\alpha_1\cdots\alpha_j}\;,
\end{eqnarray}
where the decay constant $f_{jPi}=f_{Pj_l}$ has the same value for the two states in the same doublet and $\eta^{\alpha_1\cdots\alpha_j}$ is the transverse, symmetric and traceless polarization tensor. 

In the framework of QCD sum rules the decay constant can be obtained from the two-point correlator:
\begin{eqnarray}
\label{sum1}
\Pi^{\alpha_1\cdots\alpha_j,\beta_1\cdots\beta_j}_{j,P,i}(k)=i\int d^4xe^{ik\cdot x}\langle 0|T\left(J_{j,P,i}^{\alpha_1\cdots\alpha_j}(x)J_{j,P,i}^{\dag\beta_1\cdots\beta_{j}}(0)\right )|0\rangle\;,
\end{eqnarray}
where $k$ is the residual momentum, corresponding to the decomposition $p^{\mu}=m_Qv^{\mu}+k^{\mu}$ of the heavy meson momentum.
This expression has the following form at the leading order \cite{huang} 
\begin{eqnarray}
\label{sum2}
\Pi^{\alpha_1\cdots\alpha_j,\beta_1\cdots\beta_j}_{j,P,i}(k)=(-1)^j\;\text{\bf S}\;g_t^{\alpha_1\beta_1}\cdots g_t^{\alpha_j\beta_j}\:\Pi_{j,P,i}(\omega)\;,
\end{eqnarray}
where $\omega=2v\cdot k$ is twice the external off-shell energy and $\text{\bf S}$ denotes symmetrization and subtracting the trace terms in the sets $(\alpha_1\cdots\alpha_j)$ and $(\beta_1\cdots\beta_{j})$, and $\Pi_{j,P,i}(\omega)=\Pi_{P,j_l}(\omega)$ is independent of $j$ and satisfies dispersion relations of the form:
\begin{eqnarray}
\label{dispersion}
\Pi_{j,P,i}(\omega)=\int d\nu{\rho_{j,P,i}(\nu)\over\nu-\omega}+\text{substractions}\;,
\end{eqnarray}
where $\rho_{j,P,i}(\nu)=\displaystyle{\frac{1}{\pi}\;\text{Im}\Pi_{j,P,i}(\nu)}$ represents the spectral density.

We shall confine our study to the doublets $(0^+,1^+)$ and $(1^+,2^+)$ here.
According to (\ref{current1}), (\ref{current2}) and (\ref{gamma2}), there are two possible choices for currents creating $0^+$ and $1^+$ of the doublet $(0^+,1^+)$, either
\begin{eqnarray}
\label{curr1}
J^{\dag}_{0,+,2}&=&\frac{1}{\sqrt{2}}\:\bar h_vq\;,\\
\label{curr2}
J^{\dag\alpha}_{1,+,2}&=&\frac{1}{\sqrt{2}}\:\bar h_v\gamma^5\gamma^{\alpha}_tq\;,
\end{eqnarray}
or
\begin{eqnarray}
\label{curr3}
J^{'\dag}_{0,+,2}&=&\frac{1}{\sqrt{2}}\:\bar h_v(-i)\not\!{\cal D}_tq\;,\\
\label{curr4}
J^{'\dag}_{1,+,2}&=&\frac{1}{\sqrt{2}}\:\bar h_v\gamma^5\gamma^{\alpha}_t(-i)\not\!{\cal D}_tq\;.
\end{eqnarray}
Similarly, there are two possible choices for the currents creating $1^+$ and $2^+$ of the doublet $(1^+,2^+)$. One is \footnote{There is a typo in the 
coefficient of the last term of Eq. (25) in ref. [8], the coefficient $2$ should be ${2}/{3}$. }
\begin{eqnarray}
\label{curr5}
J^{\dag\alpha}_{1,+,1}&=&\sqrt{\frac{3}{4}}\:\bar h_v\gamma^5(-i)\left(
{\cal D}_t^{\alpha}-\frac{1}{3}\gamma_t^{\alpha}\not\!{\cal D}_t\right)q\;,\\
\label{curr6}
J^{\dag\alpha_1,\alpha_2}_{2,+,1}&=&\sqrt{\frac{1}{2}}\:\bar h_v
\frac{(-i)}{2}\left(\gamma_t^{\alpha_1}{\cal D}_t^{\alpha_2}+\gamma_t^{\alpha_2}{\cal D}_t^{\alpha_1}-\frac{2}{3}\;g_t^{\alpha_1\alpha_2}\not\!{\cal D}_t\right)q\;.
\end{eqnarray}
Another choice is obtained by adding a factor $-i\not\!{\cal D}_t$ to (\ref{curr5}) and (\ref{curr6}). Note
that, without the last term in the bracket in (\ref{curr5}) the current
would couple also to the $1^+$ state in the doublet $(0^+,1^+)$ even in
the limit of infinite $m_Q$.

Usually, the currents with the least number of derivatives are used in
the QCD sum rule approach. The sum rules with them have better convergence in the high
energy region and often have better stability. However, there is a motivation for using the
currents (\ref{curr3}), (\ref{curr4}) for the doublet $(0^+,1^+)$. In the
non-relativistic quark model, which usually gives correct ordering of 
energy levels of hadron states, the doublets $(0^+,1^+)$ and $(1^+,2^+)$
are orbital p-wave states which correspond to one derivative in the space
wave functions. In fact, as proved in \cite{huang}, the coupling constant for 
the currents (\ref{curr3}), (\ref{curr4}) and that for (\ref{curr5}), 
(\ref{curr6}) are proportional to the large components of the B-S wave functions for the doublet $(0^+,1^+)$ and $(1^+,2^+)$ respectively in the non-relativistic
approximation. Therefore, we shall consider both the currents
(\ref{curr1}), (\ref{curr2}) and (\ref{curr3}), (\ref{curr4}) for the
doublet $(0^+,1^+)$.

For the doublet $(0^+,1^+)$, when the currents ${J'}_{0,+,2}$, ${J'}_{1,+,2}$ 
in (\ref{curr3}), (\ref{curr4}) are
used the sum rule (same for the two states) after the Borel transformation is found to be
\begin{eqnarray}
\label{form1}
&&f^2e^{-2\bar\Lambda/{T}}=\frac{3}{2^6\pi^2}\int_0^{\omega_c}\omega^4e^{-\omega/{T}}d\omega-\frac{1}{2^4}\,m_0^2\,\langle\bar qq\rangle\;.
\end{eqnarray}
The corresponding formula when the current $J_{0,+,2}$ and $J_{1,+,2}$ in (\ref{curr1}) and (\ref{curr2}) are used instead of
${J'}_{0,+,2}$ and ${J'}_{1,+,2}$ is the following 
\begin{eqnarray}
\label{form2}
&&f^2e^{-2\bar\Lambda/{T}}=\frac{3}{16\pi^2}\int_0^{\omega_c}\omega^2e^{-\omega/{T}}d\omega+\frac{1}{2}\,\langle\bar qq\rangle-{1\over 8T^2}\,m_0^2\,\langle\bar qq\rangle\;.
\end{eqnarray}
When the currents (\ref{curr5}) and (\ref{curr6}) are used the sum rule for
the $(1^+,2^+)$  doublet is found to be
\begin{eqnarray}
\label{form3}
&&f^2e^{-2\bar\Lambda/{T}}={1\over 2^6\pi^2}\int_0^{\omega_c}\omega^4e^{-\omega/{T}}d\omega
-\frac{1}{12}\:m_0^2\:\langle\bar qq\rangle-{1\over 2^5}\langle{\alpha_s\over\pi}G^2\rangle T\;.
\end{eqnarray}
Here $m_0^2\,\langle\bar qq\rangle=\langle\bar qg\sigma_{\mu\nu}G^{\mu\nu}q\rangle$.
In the above derivations, we have confined us to terms of the lowest order in perturbation and operators of dimension less than six. Above results shall be used in the next Section.


\section{The sum rules at the ${\cal O}(1/m_Q)$ order}
\label{sec3}

From Lorentz covariance, the correlator (\ref{sum1}) still has the form (\ref{sum2}) beyond the leading order of $1/m_Q$ expansion if we confine $k$ to be a longitudinal vector. 
But now $\Pi_{j,P,j_l}(\omega)$ depends on $j$ due to the chromomagnetic interaction operator $\cal S$.
Inserting the heavy meson eigen-state of the Hamiltonian up to the order ${\cal O}(1/m_Q)$,
the pole term on the hadron side becomes
\begin{eqnarray}
\label{pole}
\Pi(\omega)_{pole}={(f+\delta f)^2\over 2(\bar\Lambda+\delta m)-\omega}={f^2\over 2\bar\Lambda-\omega}-{2\delta mf^2\over (2\bar\Lambda-\omega)^2}+{2f\delta f\over 2\bar\Lambda-\omega}\;,
\end{eqnarray}
where $\delta m$ and $\delta f$ are of the order ${\cal O}(1/m_Q)$.

To extract $\delta m$ in (\ref{pole}) we follow the approach of \cite{ball} and consider
the three-point correlation functions
\begin{eqnarray}
\label{delta}
\delta_O\Pi^{\alpha_1\cdots\alpha_j,\beta_1\cdots\beta_j}_{j,P,i}(\omega,\omega^{'})
=i^2\int d^4xd^4ye^{ik\cdot x-ik'\cdot y}\langle 0|T\left(J_{j,P,i}^{\alpha_1\cdots\alpha_j}(x)\;O(0)\;J_{j,P,i}^{\dag\beta_1\cdots\beta_{j}}(y)\right )|0\rangle\;,
\end{eqnarray}
where $O={\cal K}$ or $\cal S$. The sum $\delta_{\cal K}\Pi+\delta_{\cal S}\Pi$ is equal to the correlator (\ref{sum1}) in the order ${\cal O}(1/m_Q)$ when $\omega=\omega^{'}$. The scalar function corresponding to (\ref{delta}) can be represented as the double dispersion integral
\begin{eqnarray}
\label{delta1}
\delta_O\Pi(\omega,\omega^{'})={1\over\pi^2}\int{{\rho}_o(s,s')dsds'\over(s-\omega)(s'-\omega^{'})}\;.
\end{eqnarray}
The pole parts for them are
\begin{eqnarray}
\label{polek}
&&\delta_{\cal K}\Pi(\omega,\omega^{'})_{pole}={f^2K\over(2\bar\Lambda-\omega)(2\bar\Lambda-\omega^{'})}+{fG_{\cal K}(\omega^{'})\over 2\bar\Lambda-\omega}+{fG_{\cal K}(\omega)\over 2\bar\Lambda-\omega^{'}}\;,\\[2mm]
\label{poles}
&&\delta_{\cal S}\Pi(\omega,\omega^{'})_{pole}={d_Mf^2\Sigma\over(2\bar\Lambda-\omega)(2\bar\Lambda-\omega^{'})}+d_Mf\left[{G_{\cal S}(\omega^{'})\over 2\bar\Lambda-\omega}+{G_{\cal S}(\omega)\over 2\bar\Lambda-\omega^{'}}\right]\;,
\end{eqnarray}
where
\begin{eqnarray}
&&{K_{j,P,j_l}}=\langle j,P,j_l|{\bar h_v\,(i D_\perp)^2 h_v}|j,P,j_l\rangle\;,\\
&&{2d_M\Sigma_{j,P,j_l}}=\langle j,P,j_l|{
   \bar h_v\,g\sigma_{\mu\nu} G^{\mu\nu} h_v}|j,P,j_l\rangle\;,\\
&&d_M=d_{j,j_l},\hspace{2.5mm}d_{j_l-1/2,j_l}=2j_l+2,\hspace{2.5mm}d_{j_l+1/2,j_l}=-2j_l.
\end{eqnarray}
Let $\omega=\omega^{'}$ in (\ref{polek}) and (\ref{poles}), and compare it with (\ref{pole})
one obtains\cite{ball}
\begin{eqnarray}
\label{delm}
\delta m=-{1\over 4m_Q}(K+d_MC_{mag}\Sigma)\;.
\end{eqnarray}
The simple pole term in (\ref{polek}) and (\ref{poles}) comes from the region in which $s(s')=2\bar\Lambda$ and $s(s')$ is at the pole for a radical excited state or in the continuum. These terms are suppressed by making double Borel transformation for both $\omega$ and $\omega'$. One obtains thus the sum rules for $K$ and $\Sigma$ as
\begin{eqnarray}
\label{form4}
&&f^2K\,e^{-2\bar\Lambda/{T}}=\int_0^{\omega_c}\int_0^{\omega_c}d\omega d\omega'e^{-(\omega+\omega')/2T}\rho_{\cal K}(\omega,\omega')\;,\\
&&f^2\Sigma\,e^{-2\bar\Lambda/{T}}=\int_0^{\omega_c}\int_0^{\omega_c}d\omega d\omega'e^{-(\omega+\omega')/2T}\rho_{\cal S}(\omega,\omega')\;,
\end{eqnarray}
where the spectral densities are obtained from straitforward calculations in HQET.

Confining us to the leading order of perturbation and the operators with dimension $D\leq 5$ in OPE, the relevant Feynman diagrams, which contribute to $\delta_{\cal K}\Pi$ and $\delta_{\cal S}\Pi$ 
are shown in Fig. 1 and Fig. 2 respectively. The graphs missing there turn out to be vanishing in the fixed point gauge $x^{\mu}A_{\mu}(x)=0$ \cite{shifman}. The spectra functions $\rho_{\cal K}$ and $\rho_{\cal S}$ can be calculated from these diagrams. For the $j_l^P=\displaystyle{{1\over 2}^+}$ doublet, we find
\begin{mathletters}
\label{msum1d}
\begin{eqnarray}
\label{form-k1d}
&&f^2K\,e^{-2\bar\Lambda/{T}}=-{3\over 2^7\pi^2}\int_0^{\omega_c} \omega^6\;e^{-\omega/T}d\omega+{3\over 2^4\pi}\langle\alpha_sGG\rangle\;T^3\;,\\
\label{form-s1d}
&&f^2\Sigma\,e^{-2\bar\Lambda/{T}}={1\over 48\pi}\langle\alpha_sGG\rangle \;T^3\;,
\end{eqnarray}
\end{mathletters}
when the currents (\ref{curr3}) and (\ref{curr4}) are used and
\begin{mathletters}
\label{msum1}
\begin{eqnarray}
\label{form-k1}
&&f^2K\,e^{-2\bar\Lambda/{T}}=-{3\over 2^5\pi^2}\int_0^{\omega_c} \omega^4\;e^{-\omega/T}d\omega-{1\over 2^5\pi}\langle\alpha_sGG\rangle\;T-{3\over 8}\;m_0^2\;\langle\bar qq\rangle\;,\\
\label{form-s1}
&&f^2\Sigma\,e^{-2\bar\Lambda/{T}}={1\over 24\pi}\langle\alpha_sGG\rangle \;T+{1\over 48}\;m_0^2\;\langle\bar qq\rangle\;
\end{eqnarray}
\end{mathletters}
when the currents (\ref{curr1}) and (\ref{curr2}) with no extra derivative are used. In contrast to (\ref{form-s1d}), there is a nonvanishing mixing condensate
term in (\ref{form-s1}). It arises from the diagram depicted in Fig. 2(b).
 For $j_l^P=\displaystyle{{3\over 2}^+}$ doublet, a straitforward calculation yields
\begin{mathletters}
\label{msum2}
\begin{eqnarray}
\label{form-k2}
&&f^2K\,e^{-2\bar\Lambda/{T}}=-{1\over 2^7\pi^2}\int_0^{\omega_c} \omega^6\;e^{-\omega/T}d\omega+{7\over 3\times 2^5\pi}\langle\alpha_sGG\rangle \;T^3\;,\\
\label{form-s2}
&&f^2\Sigma\,e^{-2\bar\Lambda/{T}}={1\over 72\pi}\langle\alpha_sGG\rangle \;T^3\;.
\end{eqnarray}
\end{mathletters}
Combining (\ref{msum1d}), (\ref{msum1}) and (\ref{msum2}) with (\ref{form1}), (\ref{form2}) and (\ref{form3}) we can obtain sum rules for $K$ and $\Sigma$ in the three cases.

The above results are not the whole story. The spin-symmetry violating term ${\cal S}$ not only causes splitting of masses in the same doublet, but also causes mixing of states with the same $j$, $P$ but different $j_l$. This mixing is characterized by the matrix element
\begin{eqnarray}
\label{mix1}
\langle j,P,j+\frac{1}{2}|\;\bar h_v\,g\sigma_{\mu\nu} G^{\mu\nu} h_v\;|j,P,j-\frac{1}{2}\rangle=-2\;m_{+-}(j^P)\;.
\end{eqnarray}
This quantity can be extracted from the correlator
\begin{eqnarray}
\label{mix2}
i^2\int d^4xd^4ye^{ik\cdot x-ik'\cdot y}\langle 0|T\left(J_{j,P,j+\frac{1}{2}}(x)\;\bar h_v\,{g\over 2}\sigma_{\mu\nu} G^{\mu\nu} h_v(0)\;J_{j,P,j-\frac{1}{2}}^{\dag}(y)\right )|0\rangle\;,
\end{eqnarray}
the double pole term of which is
\begin{eqnarray}
\label{mixpole}
-{2m_{+-}(j^P)\;f_{P,j+\frac{1}{2}}f_{P,j-\frac{1}{2}}\over (2\bar\Lambda_{+,j+\frac{1}{2}}-\omega)(2\bar\Lambda_{+,j-\frac{1}{2}}-\omega')}\;.
\end{eqnarray}
With the methods similar to that used above one can obtain the sum rule of $m_{+-}(j^P)$.
For two $1^+$ states we find
\begin{eqnarray}
\label{cross-m}
m_{\frac{1}{2},\frac{3}{2}}(1^+)\;f_{+\frac{1}{2}}f_{+\frac{3}{2}}
e^{-\bar\Lambda_{+\frac{1}{2}}/{T}}e^{-\bar\Lambda_{+\frac{3}{2}}/{T}}=
{\sqrt 6\over 72\pi}\;\langle\alpha_sGG\rangle \;T^3\;.
\end{eqnarray}
As an example, here we have used the currents (\ref{curr4}) and (\ref{curr5}) to do the analysis.

The masses of two $1^+$ states are obtained by diagonalizing the mass matrix
\begin{eqnarray}
\label{matrix}
\left(
\begin{array}{cc}
\bar\Lambda_{+,\frac{1}{2}}-\displaystyle{{1\over 4m_Q}}
(K_{+,\frac{1}{2}}-C_{mag}\Sigma_{+,\frac{1}{2}}) & \displaystyle{\frac{1}{4m_Q}}\;C_{mag}m_{\frac{1}{2},\frac{3}{2}}(1^+)\\[4mm]
\displaystyle{\frac{1}{4m_Q}}\;C_{mag}m_{\frac{3}{2},\frac{1}{2}}(1^+) & \bar\Lambda_{+,\frac{3}{2}}-\displaystyle{{1\over 4m_Q}}(K_{+,\frac{3}{2}}+5\;C_{mag}\Sigma_{+,\frac{3}{2}})
\end{array}\;\right)\;.
\end{eqnarray}
The correction to the masses due to the mixing is formally of the order of $0(1/m_Q^2)$.
However, due to the smallness of the difference between the two diagonal elements there is some possibility that this correction is numerically comparable to the ${\cal O}(1/m_Q)$ corrections.
Therefore we shall calculate it also.

\section{Numerical results and discussions}
\label{sec4}
We now turn to the numerical evaluation of these sum rules. 
We can eliminate the explicit dependence of the sum rules obtained in Section \ref{sec3} on  $f$
and $\bar\Lambda$ by means of the two-point function sum rules outlined in 
Section \ref{sec2}. Dividing the sum rules in (\ref{msum1d}), (\ref{msum1}) and (\ref{msum2}) by the sum rules in (\ref{form1}), (\ref{form2}) and (\ref{form3})
respectively, we obtain relevant expressions for the $K$ and $\Sigma$
as functions of the Borel parameter $T$ and the continuum threshold $\omega_c$.
For the QCD parameters entering the theoretical expressions, we use the following standard values
\begin{eqnarray}
\label{parameter}
\langle\bar qq\rangle&=&-(0.24 ~\text{GeV})^3\;,\nonumber\\
\langle\alpha_s GG\rangle&=&0.038 ~\text{GeV}^4\;,\nonumber\\
m_0^2&=&0.8 ~\text{GeV}^2\;.
\end{eqnarray}

The results for $K$ as a function of the Borel variable $T$, and for different values of the continuum
threshold $\omega_c$, are shown in Fig. 3. In particular, in Fig. 3a and 3b we depict the results of two kinds of sum rules for doublet $(0^+,1^+)$, which correspond to using two kinds of currents
(\ref{curr1}) (\ref{curr2}) and (\ref{curr3})  (\ref{curr4})  respectively. In Fig. 3c, we display the result
of sum rule associated with doublet $(1^+,2^+)$. 

Following the QCD sum rule procedure, we should check the  existence of stable regions for the results of $K$  in the Borel variable $T$ and the continuum
threshold $\omega_c$. As one can see from Fig. 3, stability starts at values of Borel parameter $T$ slightly larger than $0.7$, and stretches practically to $T\to\infty$.  
The familiar criterion that both the high-order power corrections and the contribution of the continuum
model should not be too large, generally speaking, less than $30\%$, rather strongly restricts the `working' region. Thus, the
stability at large $T$ values is not useful, since in this region the sum rules is strongly contaminated by higher resonance states. This continuum model contamination problem is quite severe also
in the case of sum rules analysis of ground state heavy meson  to $1/m_Q$. \cite{ball,neubert}. It originates from the high power dependence of the spectral densities. According to \cite{ball}, it is better to determine 
the working window by considering the stability of two-point sum rules,
and then to use the same values in the evaluation of the three-point
sum rules. The choice of so-called stability window does not necessarily
coincide with the stability plateau for the three-point sum rules. In the case 
of two-point function sum rules one finds that the working windows are $\omega_c=2.5-2.9$ GeV, $T=0.7-0.9$ GeV for the doublet $(0^+,1^+)$ and $\omega_c=2.7-3.2$ GeV, $T=0.7-0.9$ GeV for the doublet $(1^+,2^+)$ \cite{huang}.

From Fig. 3, taking the working regions of the sum rules as that for the 
two-pont function for the three cases, we obtain the set of values for $K$ as following:
\begin{eqnarray}
\label{result-k1}
K=-1.85\pm 0.30 ~~\text{GeV}^2\;,
\end{eqnarray} 
for the doublet $(0^+,1^+)$ when the currents (\ref{curr1}) (\ref{curr2}) without the derivative are used,
\begin{eqnarray}
\label{result-k2}
K=-1.90\pm 0.40 ~~\text{GeV}^2,
\end{eqnarray}
for the same doublet when the currents (\ref{curr3}), (\ref{curr4}) with
the derivative are used, and
\begin{eqnarray}
\label{result-k3}
K=-2.00\pm 0.40 ~~\text{GeV}^2,
\end{eqnarray}
for the doublet $(1^+,2^+)$.

The errors reflect the variations with the Borel parameter $T$ and the continuum
threshold $\omega_c$ within the working windows. As can be seen when comparing Fig. 3a and Fig. 3b with 
each other, $K$ value in Fig. 3b is more sensitive to the value of $\omega_c$ than that in Fig. 3a owing to the higher dimension of spectral density for the using of current with extra derivative.

In Fig. 4 we show the numerical results of the sum rules for $\Sigma$,
corresponding to the three chosen values of the continuum threshold $\omega_c$.
From these figures, we see that there is no stability plateau for the sum rules
of $\Sigma$. This is related to the absence of perturbation terms for $\Sigma$ in our approximation. However, due to the strategy mentioned above we can take the working window as that of two-point function 
sum rules like the case of $K$. The results for $\Sigma$ for the doublet
$(0^+,1^+)$ in the working regions of the sum rules are
\begin{eqnarray}
\label{result-s1}
\Sigma=0.023\pm 0.003 ~~\text{GeV}^2\;,
\end{eqnarray} 
 when the currents (\ref{curr1}) (\ref{curr2}) without the derivative are used
and
\begin{eqnarray}
\label{result-s2}
\Sigma=0.014\pm 0.003 ~~\text{GeV}^2\;,
\end{eqnarray}
when the currents (\ref{curr3}), (\ref{curr4}) with
the derivative are used. For the doublet $(1^+,2^+)$, the value of $\Sigma$
is given by
\begin{eqnarray}
\label{result-s3}
\Sigma=0.020\pm 0.003 ~~\text{GeV}^2\;.
\end{eqnarray}

Dividing the sum rule in (\ref{cross-m}) by the sum rules in (\ref{form2}) and
(\ref{form3}), we also obtain the expression for  $m_{\frac{1}{2},\frac{3}{2}}(1^+)$. In Fig. 5 we plot the result of this sum rules as a function of $T$, and for three values of $\omega_c$. Using the working windows as $\omega_c=2.6-3.0$ GeV and $T=0.7-0.9$ GeV, we obtain
\begin{eqnarray}
\label{result-m}
m_{\frac{1}{2},\frac{3}{2}}(1^+)=0.032\pm 0.004 ~~\text{GeV}^2\;.
\end{eqnarray}
After combining with (\ref{matrix}), we found that the correction to the masses
due to the mixing is negligibly small.

In conclusion, we have calculated the $1/m_Q$ correction to the masses of 
lowest excited heavy mesons from sum rules within the framework of the HQET.
 This study
refines the leading order analysis \cite{huang}.
From (42)-(47), together with (\ref{delm}), we can obtain the corrections to the masses of excited $D$ mesons. For the doublet $(0^+,1^+)$, our final results read
\begin{eqnarray}
\label{final-1}
{1\over 4}\;(m_{D_0^*}+3m_{D'_1})&=&m_c+\bar\Lambda+{1\over m_c}\;[(0.46\pm 0.08)~\text{GeV}^2]\;,\\
m_{D'_1}-m_{D_0^*}&=&{1\over m_c}\;[(0.023\pm 0.003)~\text{GeV}^2]\;,
\end{eqnarray}
when the currents (\ref{curr1}) and (\ref{curr2}) are used and
\begin{eqnarray}
\label{final-1d}
{1\over 4}\;(m_{D_0^*}+3m_{D'_1})&=&m_c+\bar\Lambda+{1\over m_c}\;[(0.48\pm 0.10)~\text{GeV}^2]\;,\\
m_{D'_1}-m_{D_0^*}&=&{1\over m_c}\;[(0.014\pm 0.003)~\text{GeV}^2]\;,
\end{eqnarray}
 when the currents (\ref{curr3}) and (\ref{curr4}) are used. As for the doublet
$(1^+,2^+)$, the result is
\begin{eqnarray}
\label{final-2}
{1\over 8}\;(3m_{D_1}+5m_{D_2^*})&=&m_c+\bar\Lambda+{1\over m_c}\;[(0.50\pm 0.10)~\text{GeV}^2]\;,\\
m_{D_2^*}-m_{D_1}&=&{1\over m_c}\;[(0.040\pm 0.006)~\text{GeV}^2]\;,
\end{eqnarray}
where the values of $\bar\Lambda$ in three cases are given in \cite{huang}. 
Note that we have reasonably neglected the renormalization coefficient $C_{mag}$ of the chromomagnetic operator for charmed meson in the discussion. We should also note that the given errors  present only the uncertainties of $T$ and $\omega_c$ in the working regions.

The above results for the spin splittings are not theoretically reliable because
we have not included the Feynman diagrams which are of the one loop order
in the perturbation theory and may be of values comparable to that of the
gluon condensate term leading in our approximation. Nevertheless, they are in 
approximate agreement with the experimental values for the mass splitting between  $D_1$ and $D_2^*$ \cite{review} if we take $m_c\simeq 1.3$ GeV.

\acknowledgments

This work is supported in part by the National Natural Science Foundation of China.

\newpage
{\bf Figure Captions}
\vspace{2ex}
\begin{center}
\begin{minipage}{120mm}
{\sf Fig. 1.} \small{Feynman diagrams contributing to the sum rules for $K$, in the coordinate gauge. The gray square corresponds to the insertion of kinetic energy operator at ${\cal O}(1/M_Q)$ in the HQET Lagrangian. The right vertex is put to the origin in coordinate space. There are no figures (c) (d) (h) when the currents (\ref{curr1}) and (\ref{curr2}) are used.}
\end{minipage}
\end{center}
\begin{center}
\begin{minipage}{120mm}
{\sf Fig. 2.} \small{Feynman diagrams contributing to the sum rules for $\Sigma$, in the coordinate gauge. The gray square corresponds to the insertion of chromomagnetic interaction operator at ${\cal O}(1/M_Q)$ in the HQET Lagrangian.}
\end{minipage}
\end{center}
\begin{center}
\begin{minipage}{120mm}
{\sf Fig. 3.} \small{The sum rules for $K$ as a functions of the Borel parameter
 $T$ for different values of the continuum threshold $\omega_c$. (a) (b) for $(0^+, 1^+)$ doublet with the currents (\ref{curr1}) (\ref{curr2}) and (\ref{curr3})
(\ref{curr4}) used respectively. From top to bottom the curves correspond 
to $\omega_c=2.9, 2.7, 2.5$ GeV respectively. (c) for $(1^+, 2^+)$ doublet. From top to bottom the curves correspond 
to $\omega_c=3.1, 2.9, 2.7$ GeV respectively.}
\end{minipage}
\end{center}
\begin{center}
\begin{minipage}{120mm}
{\sf Fig. 4.} \small{The sum rules for $\Sigma$ as a functions of the Borel parameter
 $T$ for different values of the continuum threshold $\omega_c$. (a) (b) for $(0^+, 1^+)$ doublet with the currents (\ref{curr1}) (\ref{curr2}) and (\ref{curr3})
(\ref{curr4}) used respectively. From top to bottom the curves correspond 
to $\omega_c=2.9, 2.7, 2.5$ GeV respectively. (c) for $(1^+, 2^+)$ doublet. From top to bottom the curves correspond 
to $\omega_c=3.1, 2.9, 2.7$ GeV respectively.}
\end{minipage}
\end{center}
\begin{center}
\begin{minipage}{120mm}
{\sf Fig. 5.} \small{The sum rules for $m_{\frac{1}{2},\frac{3}{2}}(1^+)$ as a functions of the Borel parameter
 $T$ for different values of the continuum threshold $\omega_c$ (from top to bottom: $\omega_c=3.0,2.8,2.6$). }
\end{minipage}
\end{center}


\begin{thebibliography}{1}

\bibitem{eichten}  E. Eichten, C. T. Hill and C. Quigg, Fermilab-Conf-94/118-T (1994).

\bibitem{grinstein}  B. Grinstein, Nucl. Phys. {\bf B339}, 253(1990);
                    E. Eichten and B. Hill, Phys. Lett.  {\bf B234}, 511(1990);
		    A. F. Falk, H. Georgi, B. Grinstein and M. B. Wise, Nucl. Phys. {\bf B343}, 1(1990); F. Hussain, J. G. K\"{o}rner, K. Schilcher, G. Thompson and Y. L. Wu, Phys. Lett.  {\bf B249}, 295(1990); J. G. K\"{o}rner
and G. Thompson, Phys. Lett.  {\bf B264}, 185(1991).

\bibitem{dai1}  Y. B. Dai, C. S. Huang and H. Y. Jin, Phys. Lett.  {\bf B331}, 174(1994).          

\bibitem{dai2} Y. B. Dai and H. Y. Jin, Phys. Rev. {\bf D 52}, 236(1995). 

\bibitem{bagen}  E. Bagen, P. Ball, V. M. Braun and H. G. Dosch, Phys. Lett.  {\bf B278}, 457(1992); M. Neubert, Phys. Rev. {\bf D 45}, 2451(1992); D. J. Broadhurst and A. G. Grozin, Phys. Lett.  {\bf B274}, 421(1992); C. A. Dominguez and N. Paver, Phys. Lett.  {\bf B276}, 179(1992); M. Neubert, Phys. Rep. {\bf 245}, 259(1994).

\bibitem{ball}  P. Ball and V. M. Braun, Phys. Rev. {\bf D 49}, 2472(1994).

\bibitem{neubert}  M. Neubert, Phys. Lett. {\bf B389}, 727(1996).


\bibitem{huang} Y. B. Dai, C. S. Huang, M. Q. Huang and C. Liu,  Phys. Lett. {\bf B390}, 350(1997).

\bibitem{colangelo}  P. Colangelo, G. Nardulli, A. A. Ovchinnikov and N. Paver, Phys. Lett.  {\bf B269}, 204(1991); P. Colangelo, G. Nardulli and N. Paver, Phys. Lett.  {\bf B293}, 207(1992).

\bibitem{dai3}  Y. B. Dai, C. S. Huang and H. Y. Jin, Zeit. Phys. {\bf C60}, 527(1993).

\bibitem{shifman} For a review, see V. A. Novikov, M. A. Shifman, A. I. Vainshtein and V. I. Zakharov, Fortschr. Phys. {\bf 32} 585(1984).

\bibitem{review} Particle Data Group, R. M. Barnett $et.~ al.$, Phys. Rev. {\bf D 54}, 1(1996).

\end{thebibliography}
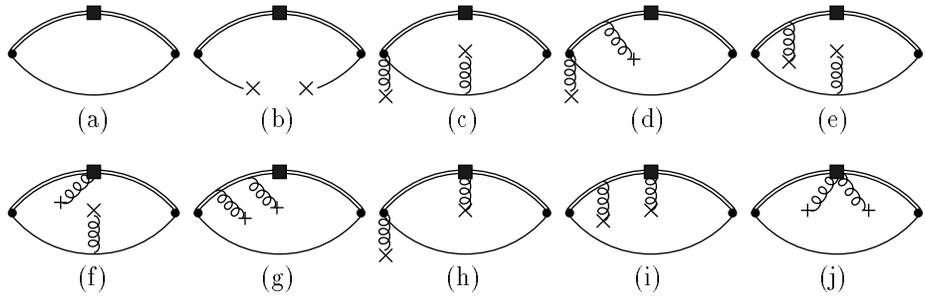
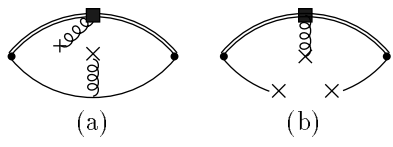
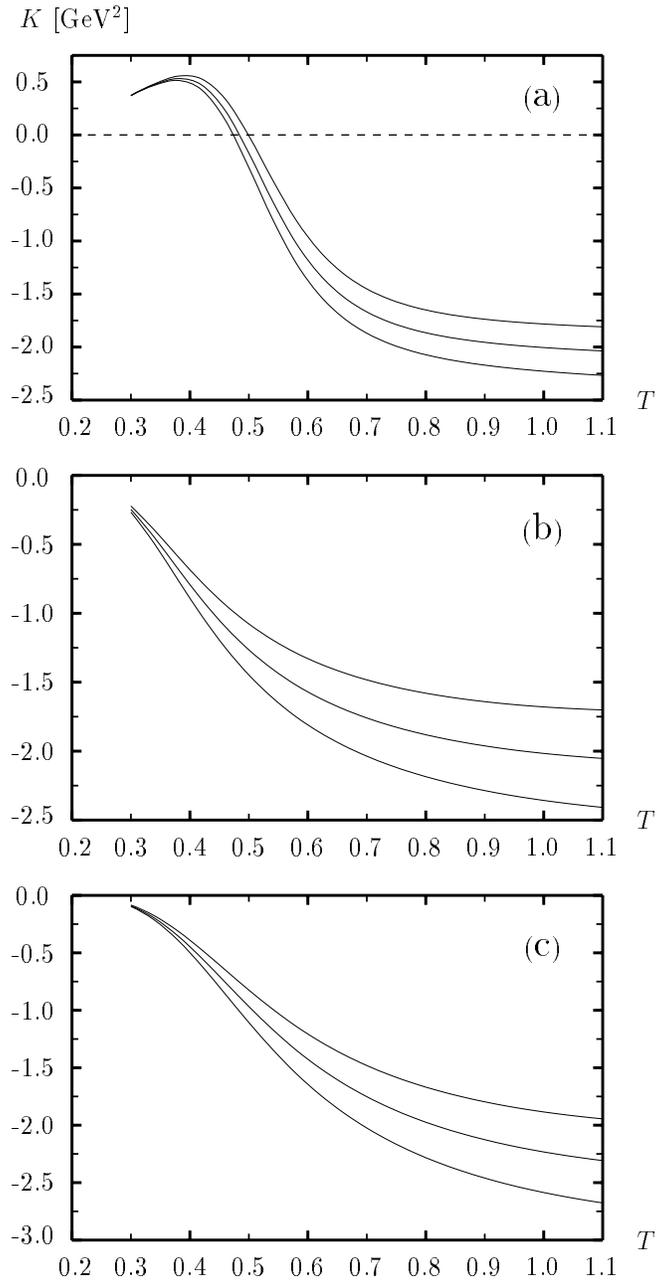
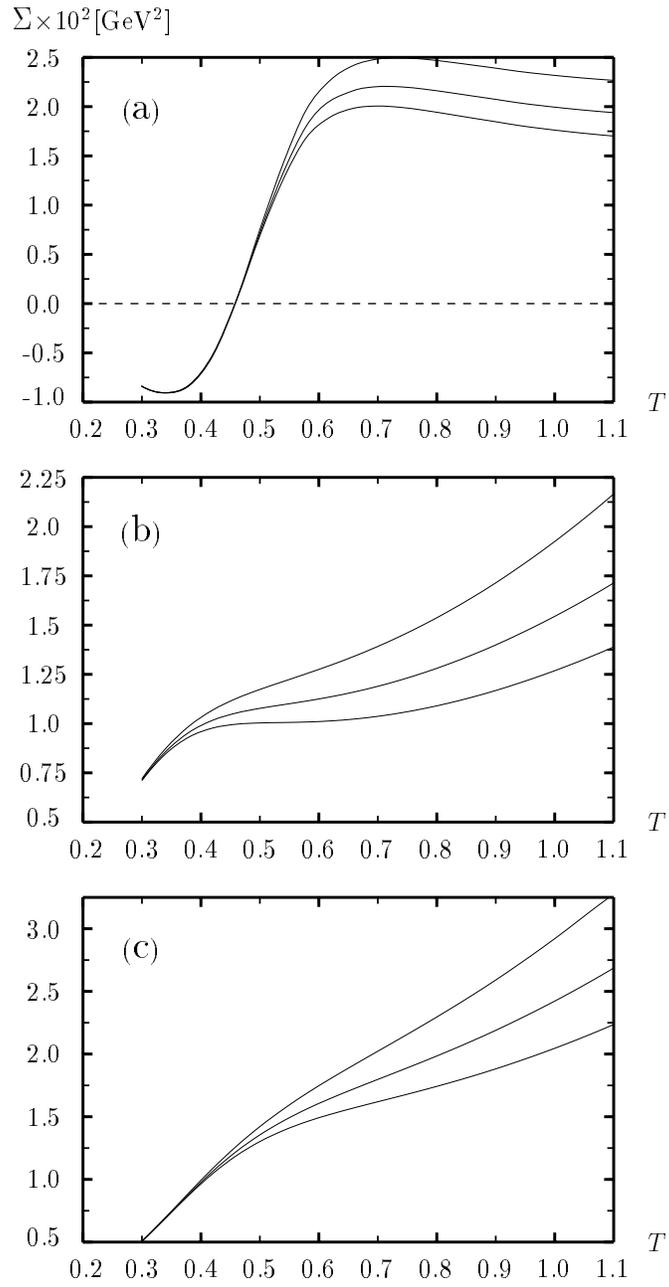
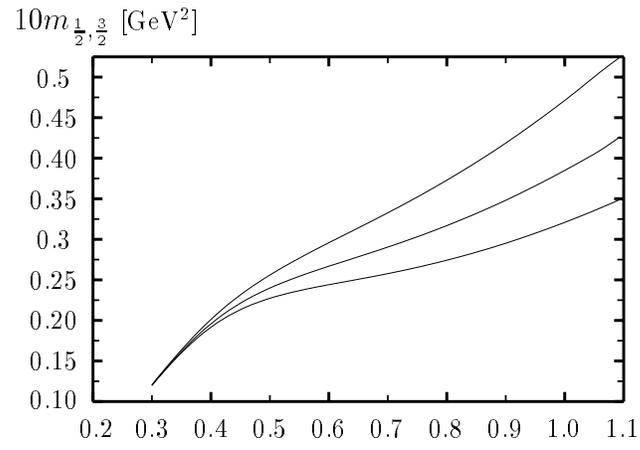
\end{document}